\documentclass[aps,prl,superscriptaddress,showpacs,preprintnumbers,twocolumn,10pt]{revtex4}
\usepackage{graphicx,color}
\newcommand{\be}{\begin{equation}}
\newcommand{\ee}{\end{equation}}
\newcommand{\bq}{\begin{eqnarray}}
\newcommand{\eq}{\end{eqnarray}}

\newcommand{\Strut}{\rule[-1.7ex]{0pt}{4.7ex}}        % a big strut
\def\lsim{\mathrel{\rlap{\lower4pt\hbox{\hskip1pt$\sim$}}\raise1pt\hbox{$<$}}}
\def\gsim{\mathrel{\rlap{\lower4pt\hbox{\hskip1pt$\sim$}}\raise1pt\hbox{$>$}}}
\def\nostrocostruttino#1\over#2{\mathrel{\mathop{\kern 0pt \rlap
{\hbox{$#1$}}} \hbox{\kern-.135em $#2$}}}

\def\Vec#1{{\bf #1}}

\def\D{{\mathrm d}}

\def\I{{\mathrm i}}%
\begin{document}
%%%%%%%%%%%%%%%%%%%%%%%%%%%%%%%%%%%%%%%%%%%%%%%%%%%%%%%%%%%%%%%%%%%%%%%%%%%%%%
%\preprint{}
\title{A leading-twist beam-spin asymmetry \\ 
in double-hadron inclusive lepto-production}

\author{Mauro Anselmino}
\affiliation{Dipartimento di Fisica Teorica, Universit{\`a} di Torino; \\
INFN, Sezione di Torino, 10125 Torino, Italy}

\author{Vincenzo Barone}
\affiliation{Di.S.T.A., Universit\`a del Piemonte Orientale
``A. Avogadro''; \\ INFN, Gruppo Collegato di Alessandria, 15121
Alessandria, Italy}

\author{Aram Kotzinian}
\affiliation{Dipartimento di Fisica Teorica, Universit{\`a} di Torino; \\
INFN, Sezione di Torino, 10125 Torino, Italy}
\affiliation{Yerevan Physics Institute, 375036 Yerevan, Armenia}

%\date{}

\begin{abstract}

We show that a new beam-spin asymmetry appears in deep inelastic 
inclusive lepto-production at low transverse momenta when a hadron 
in the target fragmentation region is observed in association 
with another hadron in the current fragmentation region. 
The beam leptons are longitudinally polarized while the target 
nucleons are unpolarized. This asymmetry is a leading-twist effect 
generated by the correlation between the transverse momentum of quarks 
and the transverse momentum of the hadron emitted by the target. 
Experimental signatures of this effect are discussed.

\end{abstract}


\maketitle

\vspace{0.5cm}

Polarization phenomena in inclusive lepto-production have recently attracted 
a huge amount of theoretical and experimental interest; new theoretical 
concepts have been introduced to interpret the data and several dedicated experiments are either running, being built or designed 
\cite{Barone:2001sp,Barone:2010zz}.  

According to the usual QCD partonic interpretation the high energy leptons 
interact with the nucleon ($N$) constituents by exchanging a virtual photon 
($\gamma^*$); the final lepton is detected, while the target and the struck 
quark fragment into final hadrons which may or may not be observed. 
A rich amount of information can be inferred from the azimuthal distribution, around the $\gamma^*$-direction, of the final particles and, eventually, 
of the nucleon transverse spin. 

In particular it has been discovered that correlations between the spin 
and the transverse momentum of quarks and/or hadrons give rise to various 
single-spin and azimuthal asymmetries that would be vanishing or negligible 
if generated by perturbative effects only. Thus, this partonic interpretation,
which holds at leading order in $1/Q$ (leading-twist), yields non perturbative 
information on the 3-dimensional spin and momentum distribution of quarks 
inside the protons and neutrons.      

One of most interesting observable is the beam-spin asymmetry 
in semi-inclusive deeply inelastic scattering (SIDIS) with 
longitudinally  polarised leptons and an unpolarized target, 
$ l^{\uparrow} \, N \to l' \, h \, X$. When the hadron is produced in the  
current fragmentation region (that is generated by the fragmentation
of a struck quark), this asymmetry, usually denoted by $A_{LU}$, is 
characterized by a $\sin \phi$ modulation, where $\phi$ is the azimuthal 
angle of the hadron measured with respect to the lepton scattering plane. 

Perturbatively, $A_{LU}$ is a $\alpha_s^2$
effect \cite{Hagiwara:1983} and  its magnitude has been 
estimated to be of order of one per mille \cite{Ahmed:1999}. 
On the other hand,  experimental studies  
\cite{Avakian:2004,Airapetian:2007,Aghasyan:2011} have shown 
that  $A_{LU}$ is much larger, of the order of few percents. 
This can be explained by taking into account 
the intrinsic transverse motion of quarks. 
In fact, transverse momentum dependent 
distribution and fragmentation functions, which reflect 
spin-orbit correlations inside hadrons, contribute  
to $A_{LU}$ at twist-3, i.e. at ${\mathcal O}(1/Q)$ level
\cite{Mulders:1995dh,Bacchetta:2006tn}. Predictions of these effects 
are difficult, as they involve 
many unknown quantities, but various simple models are 
able to explain the observed size of $A_{LU}$ \cite{Yuan:2004,Afanasev:2006}. 
The peculiarity of $A_{LU}$ 
is that it is the only observable that is expected to vanish 
in the parton model, being entirely due to 
quark-gluon interactions.  

The purpose of this Letter is to point out that, if 
a second hadron is detected in the target fragmentation region, 
a new beam-spin asymmetry arises, which is a {\it leading-twist} 
observable. This asymmetry is related to a transverse momentum 
dependent fracture function \cite{Anselmino:2011ss, Anselmino:2011bb} 
describing longitudinally polarized quarks inside an unpolarized nucleon. 
The fracture functions give the conditional probabilities to find 
a quark inside a a nucleon fragmenting into a final hadron. Upon
integration and sum over the final hadron degrees of freedom they 
reproduce the usual Transverse Momentum Dependent distribution 
functions (TMDs). The asymmetry turns out to have a typical 
azimuthal modulation as a function of $\Delta \phi$, where 
$\Delta \phi = \phi_1 - \phi_2$ is the difference between the 
azimuthal angles of the hadrons in the current and target 
fragmentation regions. 

The process we are interested in is two-particle inclusive 
lepto-production with one hadron ($h_1$) in the current fragmentation 
region (CFR) and one hadron ($h_2$) in the target fragmentation region
(TFR),
\be
 l^{\uparrow}(\ell) + p (P) \to l (\ell') + 
h_1 (P_1) + h_2 (P_2) +  X.
\ee 
We suppose that the incoming lepton is 
longitudinally polarized, while the nucleon target  
and both final hadrons are unpolarized. 

Single-particle SIDIS is usually described in terms of the three variables 
\begin{equation}
  x_B = \frac{Q^2}{2P{\cdot}q} \quad \quad
  y = \frac{P{\cdot}q}{P{\cdot}\ell} \quad \quad
  z_1 = \frac{P{\cdot}P_1}{P{\cdot}q} \> \cdot
\label{variables}
\end{equation}
When a second hadron, $h_2$, is produced,  
one needs further variables related to $P_2$. 
It is convenient to use a light-cone parametrization of vectors. 
Given a generic vector $A^{\mu} = (A^0, A^1, A^2, A^2)$,  
their light-cone components are defined as $A^{\pm} \equiv 
(A^0 \pm A^3)/\sqrt{2}$ and we write $A^{\mu} = [A^+, A^-, \Vec A_{\perp}]$. 
 We also introduce two null vectors, $n_+^{\mu} = [1, 0, \Vec 0_{\perp}]$ and 
$n_-^{\mu} = [0, 1, \Vec 0_{\perp}]$, 
with $n_+ \cdot n_- = 1$,  so that a vector 
can be parametrized as $A^{\mu} = A^+ n_+^{\mu} 
+ A^- n_-^{\mu} + A_{\perp}^{\mu}$. 

We work in a frame 
where the target nucleon and the virtual photon 
are collinear (we call it a ``$\gamma^* N$ collinear frame''). 
The nucleon is supposed to move in the negative $z$ direction. 
The unit vector $\hat{\Vec q}
\equiv \Vec q /\vert \Vec q \vert$ identifies the positive
$z$ direction. 

In terms of the null vectors $n_+^{\mu}$ and $n_-^{\mu}$
the  four-momenta at hand are (we neglect ${\mathcal O}(1/Q^2)$ terms):
\bq
& &  P^{\mu} =  P^- n_-^{\mu},  \\
& &  q^{\mu} = \frac{Q^2}{2 x_B P^-} \, n_+^{\mu} - x_B \, P^- n_-^{\mu}, \\
& & P_1^{\mu}  = \frac{z_1 Q^2}{2 x_B P^-} \, 
n_+^{\mu} + P_{1 \perp}^{\mu},    
\\
& & P_2^{\mu} =  \zeta_2 \, P^- n_-^{\mu} + P_{2 \perp}^{\mu}.
\eq

In terms of the variables  
$(z_1, \Vec P_{1 \perp})$ and $(\zeta_2, \Vec P_{2 \perp})$,  
the cross section takes the form \cite{Anselmino:2011bb}
\bq
& &  \frac{\D \sigma}{\D x_B \, \D y \, \D z_1 \,\D \zeta_2 \, 
 \D^2 \Vec P_{1 \perp} \, \D^2 \Vec P_{2 \perp}}
\nonumber \\
& & \hspace{1cm}  =
  \frac{\alpha_{\rm em}^2}{8 \, (2 \pi)^2 \,  Q^4} \, \frac{y}{z_1 \zeta_2} \,
L_{\mu\nu} W^{\mu\nu} \,.
  \label{sidis9}
\eq

\begin{figure}[t]
\begin{center}
\includegraphics[width=0.50\textwidth]
{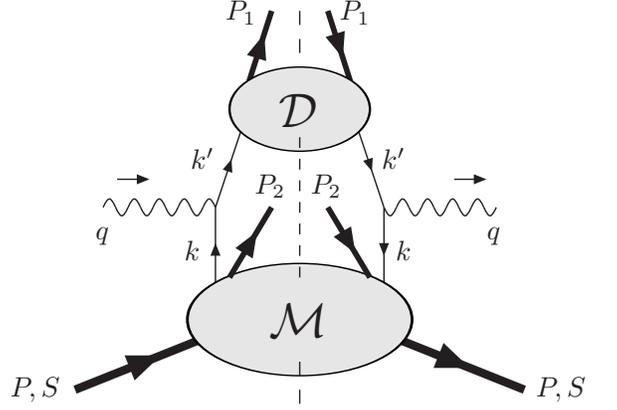}
\caption{The handbag diagram for double-hadron 
 leptoproduction. The lower blob contains the fracture functions 
 and the upper one the fragmentation functions.}
\label{handbag_two}
\end{center}
\end{figure}

$L^{\mu \nu}$ is the usual DIS leptonic tensor, explicitly given in 
\cite{Mulders:1995dh} (it can also be found in \cite{Anselmino:2011bb}). 
The hadronic tensor $W^{\mu \nu}$ is represented in the parton model, 
or, equivalently, at lowest order in QCD, by the handbag diagram of 
Fig.~1. It incorporates the transverse momentum 
dependent fracture functions introduced in \cite{Anselmino:2011ss}, which 
represent the conditional probability to find a quark with longitudinal 
momentum fraction $x_B$ and transverse momentum $\Vec k_{\perp}$ inside a 
nucleon that fragments into a hadron carrying a fraction $\zeta_2$ of the 
nucleon longitudinal momentum and a transverse momentum $\Vec P_{2 \perp}$.
The hadronic tensor also incorporates, in this case of double production, 
the fragmentation function describing the production of the final hadron 
in the CFR. In particular, for an unpolarized target, 
$W^{\mu \nu}$ explicitly reads at leading twist \cite{Anselmino:2011bb}
\bq
& & W^{\mu \nu} = 4 z_1 \zeta_2 \, (2 \pi)^3 \, \sum_a e_a^2 
\nonumber \\
& & \hspace{0.1cm} \times 
\, \int \D^2 \Vec k_{\perp} \, \int \D^2 \Vec k_{\perp}' \, 
\delta^2 (\Vec k_{\perp} - \Vec k_{\perp}' - \Vec P_{1 \perp}/z_1) 
\nonumber \\ 
& & \hspace{0.2cm} \times 
\left \{ \Strut - g_{\perp}^{\mu \nu} \hat{u}_1 \, D_1 
\right. 
\nonumber \\
& & \hspace{0.3cm} 
+ \frac{P_{2 \perp}^{\{ \mu} {k_{\perp}'}^{ \!\nu \}}- 
g_{\perp}^{\mu \nu} P_{2 \perp} \cdot k_{\perp}'}{m_1 m_2} 
\, \hat{t}_1^h \, H_1^{\perp} 
\nonumber \\
& & \hspace{0.3cm} 
+  \frac{k_{\perp}^{\{ \mu} {k_{\perp}'}^{\! \nu \}}- 
g_{\perp}^{\mu \nu} k_{\perp} \cdot k_{\perp}'}{m_1 m_N} 
\, \hat{t}_1^{\perp} \, H_1^{\perp}
\nonumber \\
& & \hspace{0.3cm} 
+ \left. 
 \I \, \epsilon_{\perp}^{\mu \nu} \,  
\frac{ \Vec k_{\perp} \times \Vec P_{2 \perp}}{m_N m_2} \, 
\hat{l}_1^{\perp h} \, D_1, \right \}. 
\label{hadten_full}
\eq 
In each term on the r.h.s, a fracture function ($\hat{u}_1$, 
$\hat{t}_1^{h}$, $\hat{t}_1^{\perp}$, $\hat{l}_1^{\perp h}$),  
which depends on $x_B$, $\zeta_2$, $\Vec k_{\perp}^2$, 
$\Vec P_{2 \perp}^2$, $\Vec k_{\perp} 
\cdot \Vec P_{2 \perp}$, multiplies a fragmentation function 
($D_1$, $H_1^{\perp}$), which depends on 
$z_1$, ${\Vec k'}^{2}$. 

A few words about the nomenclature of fracture 
functions can be helpful. We denote with
 $\hat{u}$, $\hat{l}$ and $\hat{t}$ the unintegrated 
fracture functions of unpolarized, longitudinally polarized 
and transversely polarized quarks, respectively,  
with the index 1 signaling their leading-twist character  
and the subscripts $L$ and $T$ labeling the polarization of the target 
(no subscript = unpolarized, $L$ = longitudinally polarized, $T$ = transversely 
polarized). The superscripts $h$ and $\perp$ signal the presence of 
$P_{2 \perp}^i$ and $k_{\perp}^i$ accompanying factors, respectively. 

As for fragmentation functions,  $D_1$ is the transverse-momentum 
dependent unpolarized fragmentation function 
and $H_1^{\perp}$ is the Collins function, describing the fragmentation 
of transversely polarized quarks into a spinless hadron \cite{Collins:1992kk}. 

The first three components of the hadronic tensor (\ref{hadten_full}) 
give the unpolarized cross section, $\D \sigma_{UU}$. The 
fourth one gives the helicity dependent part of the cross section for a 
longitudinally polarized lepton beam, $\D \sigma_{LU}$, 
and generates a beam-spin asymmetry. Note that there 
are two terms involving the fracture functions of transversely 
polarized quarks inside the unpolarized nucleon, $\hat{t}_1^{\perp}$
and $\hat{t}_1^{h}$, which are similar to the Boer--Mulders 
function $h_1^{\perp}$ probed in the current fragmentation region 
\cite{Boer:1997nt}.  

Contracting the hadronic tensor (\ref{hadten_full}) 
with the leptonic tensor we get the differential cross section (for details 
of the calculations, see \cite{Anselmino:2011bb, Anselmino:2011ter}):
\bq
& &  \frac{\D \sigma^{l(\lambda_l) \, N \to l \, h_1 h_2 \, X}}
{\D x_B \, \D y \, \D z_1 \,\D \zeta_2 \, 
 \D \Vec P_{1 \perp}^2 \, \D \Vec P_{2 \perp}^2 \, \D \phi_1 \, \D \phi_2}
\nonumber \\
&=&  
  \frac{\pi \alpha_{\rm em}^2}{x_B y Q^2} \, 
\left \{ \left (1 - y + \frac{y^2}{2} \right ) 
{\mathcal F}_{UU}  \right. 
\nonumber \\
&+&  
\strut (1 - y) \, {\mathcal F}_{UU}^{\cos (\phi_1 + \phi_2)}
 \cos (\phi_1 +  \phi_2) 
\nonumber \\
&+& 
\strut (1 - y) \, {\mathcal F}_{UU}^{\cos (2 \phi_1)}
 \cos (2 \phi_1) 
\nonumber \\
&+& \strut 
(1 - y) \, {\mathcal F}_{UU}^{\cos (2 \phi_2)}
 \cos (2 \phi_2) 
\nonumber \\
&-&  
\left. \lambda_l \, y \left (1 - \frac{y}{2} 
\right ) \, {\mathcal F}_{LU}^{\sin (\phi_1 - \phi_2)}
\sin \Delta \phi \right \} \nonumber \\
&\equiv& \sigma_{UU} + \lambda_l \, \sigma_{LU},  
\label{cross_full}
\eq
where $\lambda_l$ is the lepton helicity and 
the structure functions ${\mathcal F} (x_B, z_1, 
\zeta_2, \Vec P_{1 \perp}^2, \Vec P_{2 \perp}^2, 
\Vec P_{1 \perp} \cdot \Vec P_{2 \perp} )$
are given, at leading twist, by: 
\bq
& & {\mathcal F}_{UU} = {\mathcal C} \left [ \hat{u}_1 D_1 \right ], 
\label{fuu1}
 \\
& & {\mathcal F}_{UU}^{\cos (\phi_1 + \phi_2)} = 
\frac{\vert \Vec P_{1 \perp} \vert 
\vert \Vec P_{2 \perp} \vert}{m_1 m_2} \, {\mathcal C}  \left [ w_1 \, \hat{t}_1^h H_1^{\perp}\right ]
\label{fuu2}
 \\
& & 
{\mathcal F}_{UU}^{\cos (2 \phi_1)} = 
\frac{\Vec P_{1 \perp}^2}{m_1 m_N} 
\, {\mathcal C} \left [w_2 \, \hat{t}_1^{\perp} H_1^{\perp} \right ]
\label{fuu3}
 \\
& & {\mathcal F}_{UU}^{\cos (2 \phi_2)} 
= 
 \frac{\Vec P_{2 \perp}^2}{m_1 m_2} 
\, {\mathcal C} \left [w_3 \, \hat{t}_1^{h} H_1^{\perp} \right ] 
\nonumber \\
& & \hspace{1.8cm} 
+ \frac{\Vec P_{2 \perp}^2}{m_1 m_N} 
\, {\mathcal C} \left [w_4 \, \hat{t}_1^{\perp} H_1^{\perp} \right ]
\label{fuu4} \\ 
& & 
{\mathcal F}_{LU}^{\sin (\phi_1 - \phi_2)} = 
\frac{\vert \Vec P_{1 \perp} \vert \vert \Vec P_{2 \perp} \vert}{m_N 
m_2} \, {\mathcal C} \left [w_5 \, \hat{l}_1^{\perp h} D_1 \right ],    
\label{ful} 
\eq
with the following 
notation for the transverse momentum convolution 
\bq
& &  {\mathcal C} \, [f (\Vec k_{\perp}, \Vec k'_{\perp}, \ldots)] 
\equiv \sum_a e_a^2 \, x_B \,  \int \D^2 \Vec k_{\perp} \, \int \D^2 \Vec k_{\perp}' 
\nonumber \\
& & \hspace{0.5cm} 
\times \, 
\delta^2 (\Vec k_{\perp} - \Vec k_{\perp}' - \Vec P_{1 \perp}/z_1) 
 \, f (\Vec k_{\perp}, \Vec k'_{\perp}, \ldots). 
\eq 
In Eqs.~(\ref{fuu2}-\ref{ful}) the $w_i$'s are scalar functions of 
the vectors $\Vec k_{\perp}$, $\Vec k_{\perp}'$, $\Vec P_{1 \perp}$, 
$\Vec P_{2 \perp}$, whose explicit expressions will be given 
in \cite{Anselmino:2011ter} (in the sequel, we will write down 
explicitly the only coefficient we need here, {\it i.e.} $w_5$).  

If we integrate the cross section over $\phi_2$,  
keeping $\Delta \phi \equiv \phi_1 - \phi_2$ fixed, the only two surviving structure functions 
in (\ref{cross_full}) are ${\mathcal F}_{UU}$ and 
${\mathcal F}_{LU}^{\sin \Delta \phi}$ (the transverse spin of quarks 
in this case plays no role).  
The beam spin asymmetry ${\mathcal A}_{LU}$, defined as  
\be
{\mathcal A}_{LU} (x_B, z_1, 
\zeta_2, \Vec P_{1 \perp}^2, \Vec P_{2 \perp}^2, 
\Delta \phi )
= \frac{\int \D \phi_2 \, \sigma_{LU}}{\int \D \phi_2 
\, \sigma_{UU}} , 
\ee
is given by 
\bq
{\mathcal A}_{LU} &=& 
- \frac{y \left (1 - \frac{y}{2} 
\right )}{\left (1 - y + \frac{y^2}{2} \right )}
\frac{{\mathcal F}_{LU}^{\sin \Delta \phi}}{{\mathcal F}_{UU}}
\sin \Delta \phi
\nonumber \\
&=& 
-  \frac{\vert \Vec P_{1 \perp} \vert \vert \Vec P_{2 \perp} \vert}{m_N 
m_2}
\frac{y \left (1 - \frac{y}{2} 
\right )}{\left (1 - y + \frac{y^2}{2} \right )}
\nonumber \\
& & \times \, 
\frac{{\mathcal C} [w_5 \, \hat{l}_1^{\perp h} D_1]}{{\mathcal C} [\hat{u}_1 D_1]} 
\sin \Delta \phi,  \label{ALU}
\eq
with
\be 
w_{5} = 
\frac{(\Vec k_{\perp} \cdot \Vec P_{2 \perp}) (\Vec P_{1 \perp} 
\cdot \Vec P_{2 \perp}) - (\Vec k_{\perp} \cdot \Vec P_{1 \perp}) 
\Vec P_{2 \perp}^2}{(\Vec P_{1 \perp} \cdot \Vec P_{2 \perp})^2 
- \Vec P_{1 \perp}^2 \Vec P_{2 \perp}^2} \cdot 
\ee

Let us consider now the most general azimuthal dependence of 
Eq. (\ref{ALU}), in order to extract a clear signature for 
the spin-beam asymmetry ${\mathcal A}_{LU}$. One should remember 
that all structure functions can carry a dependence on 
$\Vec P_{1 \perp} \cdot \Vec P_{2 \perp} = \vert \Vec P_{1 \perp} \vert 
\vert \Vec P_{2 \perp}\vert \,   \cos \Delta \phi$,  arising from a 
transverse momentum correlation of the type 
$\Vec k_{\perp} \cdot \Vec P_{2 \perp}$ in the fracture functions. 
It is reasonable to assume that these correlations are small, 
so that we can expand the fracture functions in powers of 
$\Vec k_{\perp} \cdot \Vec P_{2 \perp}$ keeping only the first few 
terms; for instance:
\bq
& & \hat{l}_1^{\perp h} (x_B, \zeta_2, \Vec k_{\perp}^2, 
\Vec P_{2 \perp}^2, \Vec k_{\perp} 
\cdot \Vec P_{2 \perp}) 
\nonumber \\ 
& & \hspace{0.5cm} \simeq 
a (x_B, \zeta_2, \Vec k_{\perp}^2, 
\Vec P_{2 \perp}^2) \nonumber \\
& & \hspace{0.7cm} + \, b (x_B, \zeta_2, \Vec k_{\perp}^2, 
\Vec P_{2 \perp}^2) \, \Vec k_{\perp} \cdot \Vec P_{2 \perp} \>.
\eq
The term linear in $\Vec k_{\perp} \cdot \Vec P_{2 \perp}$ 
yields a $\cos \Delta \phi$ term in the structure functions, which, 
combined with the $\sin \Delta \phi$ term already explicitly appearing 
in Eq.~(\ref{ALU}), results in the following angular dependence for the 
beam-spin asymmetry:
\bq
& & {\mathcal A}_{LU} (x_B, z_1, \zeta_2, \Vec P_{1 \perp}^2, 
\Vec P_{2 \perp}^2, \Delta \phi) 
\nonumber \\
& & \hspace{0.2cm} 
= A (x_B, z_1, \zeta_2, \Vec P_{1 \perp}^2, 
\Vec P_{2 \perp}^2) \, \sin \Delta \phi 
\nonumber \\ 
& & \hspace{0.4cm} + \, 
B (x_B, z_1, \zeta_2, \Vec P_{1 \perp}^2, 
\Vec P_{2 \perp}^2) \, \sin (2 \Delta \phi).   
\eq

These two azimuthal modulations are typical of ${\mathcal A}_{LU}$ and
would be a clear signature of its presence; they both originate from a
correlation between the quark transverse momentum $\Vec k_\perp$ and 
the hadron transverse momentum $\Vec P_{2 \perp}$, resulting in a 
long range correlation between $\Vec P_{1 \perp}$, the momentum of the
hadron in the CFR, and $\Vec P_{2 \perp}$, the momentum of the hadron 
in the TFR, which yields a specific and unambiguous dependence on 
$\phi_1 - \phi_2$. 

We have shown that a new beam-spin asymmetry appears, at leading 
twist and low transverse momenta, in the deep inelastic inclusive 
lepto-production of two hadrons, one in the target fragmentation region 
and one in the current fragmentation region. Such a single spin 
asymmetry, if sizable, cannot originate from pQCD effects and must be 
related to non perturbative properties of partonic distributions. 
When interpreted in the QCD parton model and in a non collinear 
factorization scheme (TMD factorization) based on fracture and 
fragmentation functions, the asymmetry has a clear partonic origin in 
a correlation between the quark intrinsic motion and the transverse motion 
of the hadron produced in the TFR. The TMD factorization has been widely
used in the CFR to obtain information on TMDs and the 3-dimensional 
momentum structure of the nucleon, and this is known to hold in the TFR 
as well \cite{Grazzini:1997ih,Collins:1997sr,Collins:2011}
% Berera:1994xh} 

The new beam-spin asymmetry introduced here has a definite and clear 
signature which can be experimentally tested, both in running experiments 
(JLab) and future ones (upgraded JLab and future electron-ion or 
electron-nucleon colliders, EIC/ENC). If experimentally observed, it would 
confirm the validity of the TMD factorization in high energy lepto-production
for TFR events, thus opening new ways of exploring the nucleon internal 
structure.

\begin{acknowledgments}
We acknowledge support by the European Community - Research Infrastructure
Activity under the FP6 Program ``Structuring the European Research Area''
(HadronPhysics, contract number RII3-CT-2004-506078), by the Italian 
Ministry of Education, University and Research (PRIN 2008) and  
by the Helmholtz Association through
funds provided to the virtual institute ``Spin and Strong QCD''(VH-VI-231).
The work of one of us (A.K.) is also supported by 
Regione Piemonte. 

\end{acknowledgments}

\bibliographystyle{h-physrev3.bst}

%\bibliography{biblio_costwophi}

\end{document}